\documentclass[aps,twocolumn,prb,superscriptaddress,amsmath,showpacs,tightenlines]{revtex4}\usepackage[dvipdfmx]{graphicx}
\usepackage{graphicx}
\usepackage{epsfig}
\usepackage{amssymb}

\newcommand{\dd}{\downarrow}
\newcommand{\uu}{\uparrow}

\begin{document}
\title{
Generation of chiral spin state by  quantum simulation}
%
\author{Tetsufumi Tanamoto}
\affiliation{Corporate R \& D center, Toshiba Corporation,
Saiwai-ku, Kawasaki 212-8582, Japan}

\date{\today}
\begin{abstract}
Chirality of materials in nature appears when there are asymmetries in their lattice structures or interactions in a certain environment.  
Recent development of quantum simulation technology has enabled the manipulation of qubits.  
Accordingly, chirality can be realized intentionally rather than passively observed.
Here we theoretically provide simple methods to create a chiral spin state in a spin-1/2 qubit system on a square lattice.
First, we show that switching ON/OFF 
the Heisenberg  and $XY$ interactions 
produces the chiral interaction directly in the effective Hamiltonian without controlling local fields.
Moreover, when initial states of spin-qubits are appropriately prepared, we prove that the 
chirality with desirable phase is dynamically obtained.
Finally, even for the case where switching ON/OFF the interactions is infeasible
and the interactions are always-on, 
we show that, by preparing an asymmetric initial qubit state,  
the chirality whose phase is $\pi/2$ 
is dynamically generated.
\end{abstract}

\pacs{03.67.Lx, 03.67.Mn, 73.21.La}
\maketitle
\section{Introduction}
Chirality specifies the properties of 
materials in which 
the mirror image does not coincide with itself by 
simple rotations and translations~\cite{Rikken}. 
Recently, chirality has come to play an important role in the stabilization of skyrmions~\cite{Nagaosa,tokura},
and, in spintronic devices, chirality is observed in the domain wall motion 
through the Dzyaloshinsky-Moriya interaction~\cite{Parkin,Emori}.
When the chirality of a spin system 
supports a nonlocal extension of the order parameter, 
it is called a chiral spin liquid (CSL) that has attracted much attention in the research 
of high $T_{\rm c}$ superconductors since the 1980's~\cite{Wen,Wilczek,WWZ,Affleck,Kapitulnik}. 
The research of CSL has been developed 
in combination with topological quantum computation
~\cite{YK,Yao}.

The chiral spin state 
is represented by the chiral interaction
$\vec{S}_i \cdot \vec{S}_j \times \vec{S}_k $
($i,j,k$ indicate lattice sites)~\cite{WWZ}.
In the Hubbard model,
which can abstract the nature of strongly-correlated electrons,  
the chiral interactions appear only in the higher order 
of $t/U$-expansion, 
and are much smaller  
than the major Heisenberg couplings 
$\vec{S}_i \cdot  \vec{S}_j$~\cite{Sen}.
Numerical studies~\cite{Meng,Motrunich}
showed the spin-liquid phase appears only in the limited parameter region
of the Hubbard model. 
On the other hand, 
theoretically designed Hamiltonians
~\cite{Greiter,YK} whose ground states are the CSL are 
mathematically well-established. However,  
it is difficult to synthesize corresponding materials.

\begin{figure}
\begin{center}
\includegraphics[width=7cm]{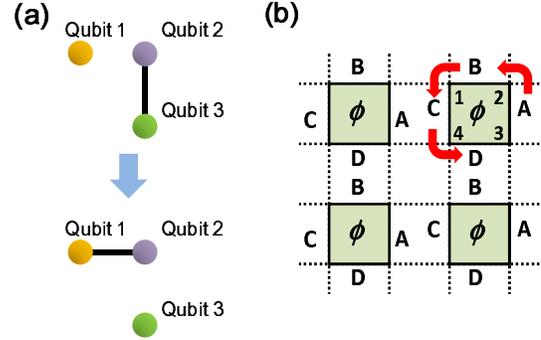}
\end{center}
\caption{
Dynamical creation of chiral interaction. 
(a) Schematic illustration of generating the chiral interaction $\hat{\chi}_{123}\equiv \vec{\sigma}_1 \cdot \vec{\sigma}_2 \times \vec{\sigma}_3 $ by switching ON/OFF 
the Heisenberg interactions between qubits.
In the first step, the Heisenberg interaction is switched ON between 
qubits 2 and 3. In the next step, after switching OFF the interaction between qubits 2 and 3,
the interaction between qubits 1 and 2 is switched ON.
Then, the effective interaction 
$\hat{\chi}_{123} $
is generated in addition to the Heisenberg couplings. 
(b),Generating process of the chiral interaction of the form of 
$(-\hat{\chi}_{123}-\hat{\chi}_{134}-\hat{\chi}_{124}+\hat{\chi}_{234}$) whose expectation value  corresponds to Eq.~(\ref{WWZ1}),
assuming that spin qubits are placed on each node of the square lattice.
The interactions between two qubits are switched ON and OFF 
in the order of A$\rightarrow$B$\rightarrow$C$\rightarrow$D, where 
A$\sim$D indicate the interactions between two qubits. }
\label{fig1}
\end{figure}

Instead of finding materials  
that have target chiral properties, 
recent quantum simulation technologies~\cite{Bloch,Franco}
can be applied to dynamically simulate the chirality of a spin system.
Here, we propose practical methods of controlling the chirality 
in a spin-qubit system on a square lattice by switching ON/OFF the 
interaction between qubits, 
or initializing the qubit states asymmetrically (spin-up or spin-down).
Typical spin-qubits are realized in semiconductor quantum dot (QD) systems
\cite{Petta,Koppens,Maune,Veldhorst}.
Each QD includes one excess electron whose spin degree of freedom 
plays the role of qubit. 
The exchange coupling $J$ is caused by Coulomb interactions between electrons, 
and is controlled by the gate electrodes. 
The switching ON/OFF of the coupling $J$ is more feasible 
than the control of the arbitrary rotation of each qubit~\cite{DiVincenzo}.
In addition, because the coherence time is limited,
the quantum operations should be as simple as possible.

In this paper, we provide three methods to create a chiral spin state in a spin-1/2 qubit system on a square lattice by switching ON/OFF of the coupling $J$.
As mentioned above, 
it is difficult to obtain the chiral interaction $\vec{S}_i \cdot \vec{S}_j \times \vec{S}_k$
as the dominant term in the conventional perturbation theory. 
In the first method, we show that 
effective chiral Hamiltonians can be designed only by 
switching ON/OFF the Heisenberg  and $XY$ interactions. 
In the second and third methods, we derive  
chiral states with arbitrary phase by preparing 
appropriate initial qubit states.
Here, we consider product states of  spin-up or spin-down
as the initial qubit states. 
This is because preparation of product states is 
much easier than that of entangled states.
Moreover, focusing on product states 
makes the discussion simple and clear.
In the second method, 
we show analytical forms of chiral states for four qubits on a square lattice.
The third method provides how to control   
chiralities by preparing appropriate initial qubit states in an always-on lattice system. 
We show that asymmetrically-arranged qubit states periodically generate
the chirality whose phase is $\pi/2$. 
In this paper, we would like to describe the clear relationship
between the phase of the chirality and the asymmetric spin state.

This paper is organized as follows:
In Sec.~\ref{sec:form} we show how to generate 
effective chiral Hamiltonians by switching ON/OFF the coupling 
between qubits.
In Sec.~\ref{sec:second}, we show our second method in 
which analytical form of the chirality is derived in a four-qubit system. 
In Sec.~\ref{sec:third}, we consider the chiral spin state 
on the always-on lattice system.
In Sec.~\ref{sec:discussions}, 
we mention experimental possibilities. 
Sec.~\ref{sec:summary} is devoted to a summary. 
In Appendix, we show detailed derivations 
of equations and related numerical calculations.

\section{Construction of effective chiral Hamiltonian}\label{sec:form}
The chirality is defined around the loop with gauge-invariant form following Ref.~\cite{WWZ}.
When $
\hat{\chi}_{ij} \equiv \sum_{s=\uparrow \downarrow} \hat{c}_{is}^\dagger \hat{c}_{js},
$
($\hat{c}_{is}$ is the electron annihilation operator),
the chirality of a square lattice is defined by
\begin{eqnarray}
w_{1234}&=&\langle \hat{\chi}_{12}\hat{\chi}_{23}\hat{\chi}_{34}\hat{\chi}_{41} \rangle.
\label{w1234}
\end{eqnarray}
In this definition, qubit 1 is the origin and end of the loop.
Thus, the asymmetry is discussed from the view of qubit 1.
The chiral spin state is defined as a state where the imaginary part of $w_{1234}$ 
has a finite phase ($w_{1234}=|w_{1234}|\exp i\phi$ and $\phi \ne 0$). 
The phase $\phi$ of the loop is proportionate to the area 
of the loop and 
is important for the topological aspect of the qubit system~\cite{Rokhsar,Kitaev,Ioffe}.
We treat a spin-1/2 model  $\vec{S}_i=(1/2)\vec{\sigma}_i$,
where $\vec{\sigma}_i =(\sigma^x_i,\sigma^y_i,\sigma^z_i)$ 
($\sigma^x_i$, $\sigma^y_i$, and $\sigma^z_i$ 
show the Pauli matrices of the lattice site $i$),
focusing on the phase of the chirality 
rather than the properties of the spin-liquid. 
The expectation value 
of $E_{123}\equiv \langle\vec{\sigma}_1 \cdot \vec{\sigma}_2 \times \vec{\sigma}_3 \rangle$
has a simple relation with the chirality of the loop~\cite{WWZ}:
For the square lattice, we have
\begin{equation}
{\rm Im} w_{1234}=
\frac{1}{8}(-E_{123}-E_{134}-E_{124}+E_{234}).
\label{WWZ1}
\end{equation}

Here, the asymmetry of the spin system is introduced by the asymmetric switching of the 
nearest-neighbor qubit-qubit interaction and the asymmetric spin configuration 
on the square lattice.  
First, we show how to obtain
the chiral interaction $\vec{\sigma}_1 \cdot \vec{\sigma}_2 \times \vec{\sigma}_3$
by switching ON/OFF 
the nearest-neighbor interactions between qubits.  
For the Heisenberg model, we use the basic relation between three spins given by
\begin{equation}
[\vec{\sigma}_1 \cdot \vec{\sigma}_2, \vec{\sigma}_2 \cdot \vec{\sigma}_3]=2i\vec{\sigma}_2 \cdot \vec{\sigma}_1 \times \vec{\sigma}_3.
\label{3product}
\end{equation}
The point is that the left commutation relation of this equation is 
obtained by simply multiplying the time-evolution operators 
$U_{ij}^{HS}(t)\equiv e^{itH^{HS}_{ij}}= \exp \{iJt \vec{\sigma}_i \cdot \vec{\sigma}_j \}$ 
in the Baker-Campbell-Hausdorf  formula given by
\begin{eqnarray}
& &
U_{12}^{HS}(t_1)U_{23}^{HS}(t_2)
=\exp\{iJ(t_1 \vec{\sigma}_1 \cdot \vec{\sigma}_2+t_2  \vec{\sigma}_2 \cdot \vec{\sigma}_3)
\nonumber \\
& &
-J^2t_1t_2/2[\vec{\sigma}_1 \cdot \vec{\sigma}_2, \vec{\sigma}_2 \cdot \vec{\sigma}_3] +...\}.
\label{eqHS} 
\end{eqnarray}
Figure 1(a) shows this process: the first step is 
switching ON the interaction between spin 2 and 3, and the next step is,  
after switching OFF this interaction,  switching ON the interaction 
between spin 1 and 2.
This process is generalized to obtain the chiral interactions 
as the next dominant terms of the effective Hamiltonian.
\begin{equation}
H_{\rm eff}=\sum_{ij} J_{ij}  \vec{\sigma}_i \cdot \vec{\sigma}_j 
+\sum_{ijk} J'_{ijk} \vec{\sigma}_i \cdot \vec{\sigma}_j \times \vec{\sigma}_k, 
\label{eqHS2}
\end{equation}
where $J'_{ijk}=J_{ij}J_{jk}t_0.$ under the condition of $J_{ij}t_0<1$ 
when $t_1=t_2=t_0$ in Eq.~(\ref{eqHS}).
As an example, the Hamiltonian whose chiral interaction has the form of  Eq.~(\ref{WWZ1}) is realized by the serial operations 
given by
$
U_{34}^{HS}(t)U_{41}^{HS}(t)U_{12}^{HS}(t)U_{23}^{HS}(t)
$.
Figure 1(b) shows this process graphically.

For the $XY$ Hamiltonian 
$H^{xy}=\sum_{i<j} H^{xy}_{ij}=
\sum_{i<j} J[\sigma_i^x \sigma_j^x+\sigma_i^y \sigma_j^y]$,  
we can generate the 
pure chiral Hamiltonian 
$H\propto \vec{\sigma}_1 \cdot \vec{\sigma}_2 \times \vec{\sigma}_3$ 
by using the equation given by
\begin{equation}
O_{ijk}^{xy}\equiv [U_{jk}^{xy}]^{-1} (\sigma^x_i\sigma^x_j+\sigma^y_i\sigma^y_j)U_{jk}^{xy}=\frac{1}{2} \sigma^z_j (\sigma^x_k\sigma^y_i-\sigma^x_i\sigma^y_j). \nonumber
\end{equation}
where  $U_{ij}^{xy}=\exp i(\pi/4) [\sigma_i^x \sigma_j^x+\sigma_i^y \sigma_j^y]$.  
The chiral Hamiltonian $H\propto \vec{\sigma}_1 \cdot \vec{\sigma}_2 \times \vec{\sigma}_3$ 
is obtained by the sequence of switching ON/OFF the $XY$ interactions:
$O_{123}^{xy}O_{231}^{xy}O_{312}^{xy}$.

\section{Construction of chiral spin state starting from product states}\label{sec:second}
The above-mentioned method is  effective
when the target Hamiltonian is not complicated.
Here, we provide a simpler method to 
obtain the chirality directly. 
When we look at the process of Fig.~1(a), it is found 
that switching ON one interaction can realize the finite chirality.
That is, the expectation value, 
$w_{1234}^{HS}(t)=\langle\Psi_0 | U_{23}^{HS\dagger} (t) \hat{\chi}(1234)
U_{23}^{HS} (t) |\Psi_0 \rangle$
with $\hat{\chi}(1234)\equiv \hat{\chi}_{12}\hat{\chi}_{23}\hat{\chi}_{34}\hat{\chi}_{41}$
and $|\Psi_0 \rangle=  |s_1s_2s_3s_4 \rangle$, 
is given by 
$
w_{1234}^{HS}(t)=(1+Z_2Z_3)(n_{1\uu}n_{4\dd}+n_{1\dd}n_{4\uu})/4
-(Z_2+Z_3+i\sin(4Jt)[Z_2-Z_3])(n_{1\uu}n_{4\dd}-n_{1\dd}n_{4\uu})/4
$,
where $Z_i=\langle s_i |\sigma_i^z |s_i \rangle$, and
$n_{is}(\in \{0,1\})$ is the number of the $s$-spin for the site $i$ 
(See the Appendix~\ref{sec:A} and ref~\cite{tanaSW,tanaSR}).
When $Z_3=-Z_2$($s_2=\uu$, $s_3=\dd$, or $s_2=\dd$, $s_3=\uu$), we have
$w_{1234}^{HS}(t)= -i \sin (4Jt)Z_2  (n_{1\uu}n_{4\dd}-n_{1\dd}n_{4\uu})/2$.
This means that  switching ON one interaction itself
generates the chirality of a phase $\pi/2$. 
The same form is obtained for the $XY$ interaction. 
The chirality of the Ising $XX$ interaction has a similar form except for 
$\sin (2Jt)$ instead of  $\sin (4Jt)$.

Thus, the chiral spin states can be dynamically created by 
directly manipulating the interactions between qubits. 
This is because the basic Eq.~(\ref{3product}) 
appears many times in the calculation of the expectation 
value $w_{1234}$.
Moreover,  switching ON two interactions 
enables the generation of the chiral state with a phase in the 
range of $-\pi$ to $\pi$, as shown in Table I.
The case on the left in Table I shows the process shown in Fig.~1(a).
The time-saving method is shown in the case on the right in Table I,
in which the interaction between 1 and 4 and that between 2 and 3 are 
simultaneously switched ON
(the general expressions are shown in the Appendix).
For example, the flux state whose phase is $\pi$~\cite{Affleck,WWZ} is given periodically 
when $4Jt=\pi/2$ for the Heisenberg interaction.
\begin{table*}
\begin{tabular}{|lr|c|c|}\hline
&  
&  Chirality of 
\begin{minipage}{1.5cm}
      \centering
      \scalebox{0.4}{\epsfig{file=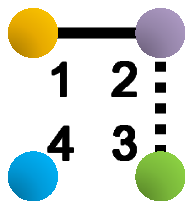,width=2.6cm}}
    \end{minipage}
&  Chirality of 
\begin{minipage}{1.5cm}
      \centering
      \scalebox{0.4}{\epsfig{file=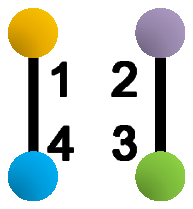,width=2.6cm}}
    \end{minipage}
\\ \hline
{\bf Ising $XX$}
&
&  
$\{e^{4iJt}-1-2i \sin (2Jt) Z_1Z_2\}/8 $ 
&
$e^{-4iJt}(Z_1 +Z_2 e^{4iJt})[Z_2-Z_1]/8$
\\
{\bf $XY$} 
& 
& 
$i e^{2iJt} \sin (4Jt) [Z_2-Z_1]^2/8 $ 
& 
$ie^{-4iJt}\sin (4Jt) [Z_2-Z_1]^2/8$
%
\\
{\bf Heisenberg}
& 
& 
$i e^{4iJt} \sin (4Jt) [Z_2-Z_1]^2/8 $ 
& 
$ie^{-4iJt}\sin (4Jt) [Z_2-Z_1]^2/8$
\\ \hline
\end{tabular}
\begin{flushleft}
Table I: Switching ON two interactions to create the chiralities with desired phases. 
The chirality on the square lattice for 
the three interactions.  For simplicity, we show the cases of 
$Z_3=Z_1=\langle s_1 |\sigma_1^z |s_1 \rangle$ and 
$Z_4=Z_2=\langle s_2 |\sigma_2^z |s_2 \rangle$ ($s_i \in \{\uparrow, \downarrow\}$).
$Z_i=1$ for $s_i=\uu$ and  $Z_i=-1$ for $s_i=\dd$. 
General form of the left switching pattern, see Appendix ~\ref{sec:tableIa} and \ref{sec:tableIb} for detail.
\end{flushleft}
\end{table*}

\begin{figure}
\begin{center}
\includegraphics[width=8cm]{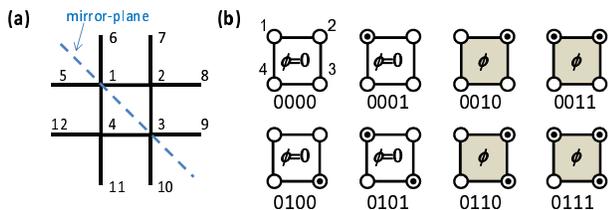}
\caption{
Configuration of spin qubits for the always-on interaction.
(a) The twelve qubit sites for numerical calculation of the chirality $w_{1234}$. The sites are connected by
the always-on interactions. The dashed line shows the mirror-plane 
when the chirality is defined by Eq.~(1).
(b), There are 2$^{4}=16$ initial states for the spin configurations
for the four qubits. 
Half of the spin configurations of the initial states are illustrated. Other configurations 
have the same results because of the symmetry. 
The circle and the double circles 
indicate the spin-up ($\uparrow$) and the spin-down ($\downarrow$) states,
respectively. Colored patterns (0010,0011,0110,0111), whose spin configurations are
asymmetric to the mirror-plane of Fig.(a),  have a phase $\pi/2$ 
at $t\sim 0$ (see Fig.~3). 
}
\end{center}
\label{square_pattern}
\end{figure}


\section{Chiral spin state with always-on interactions}\label{sec:third}
Finally, let us consider a case of more restrictive condition
in which the interactions between qubits are always-on (Fig.~2(a)).
This happens when the distances between qubits are small 
in order to reduce decoherence.
For this case, we generate the chirality only by preparing 
asymmetric initial qubit states.
Because of the commutability of the Ising interactions $[\sigma^x_i\sigma^x_j, \sigma^x_j\sigma^x_k]=0$,
the time-dependent chirality of the $XX$ interaction can be derived analytically.
On the other hand, the time-dependent chiralities of the $XY$ and the Heisenberg interactions
are obtained by numerical calculations.

The chirality of the $XX$ interaction on the square lattice   
$w_{1234}^{XX}(t)=\langle \Psi_0 | U^{xx\dagger} (t) \hat{\chi}(1234) U^{xx} (t) |\Psi_0 \rangle$
with $U^{xx} (t) \equiv \exp \{ iJt \sum_{i<j}^{1,..,12}X_i X_j \} $ and $|\Psi_0\rangle=  |s_1s_2s_3s_4 \rangle$ is given by
\begin{eqnarray}
\lefteqn { w_{1234}^{XX}(t)\!=\!\!\{ \cos^2 2Jt[ 
   Z_3 ( Z_2 e^{4iJt}  + Z_4 e^{-4iJt})
-Z_2(Z_1-Z_4)}
\nonumber \\
& &  
-Z_1(Z_3+Z_4)    
-  \cos^2 2Jt \cdot Z_1 Z_2 Z_3 Z_4 ]+1 \}/8.
\label{XX_always}
\end{eqnarray}
Note that $w_{1234}^{XX}(t)$ is irrelevant to the spin configurations of the qubits around.
Thus, when $Z_2=-Z_4$ (the colored patterns shown in Fig.~2(b)), we have 
$w_{1234}^{XX}(t) \rightarrow iZ_3 Z_2 Jt$ at $t\sim 0$, 
which means that the chirality of the Ising interaction has a phase $\pi/2$ at $t \sim 0$. 
Because of the uniform interactions between qubits,
the asymmetry is introduced by the asymmetric configuration of the qubit state
seen from qubit 1.
Figures~3(a) and (b) 
show the time-dependent amplitude and phase of $w_{1234}^{XX}$ of Eq.~(\ref{XX_always}).
Compared with the switching ON one interaction mentioned above,
we need to control the four qubit states to obtain  the $\pi/2$-phase.

Figures~3(c-f) show the numerically-calculated time-dependent chiralities of the $XY$ and the Heisenberg interactions given by 
\begin{eqnarray}
w_{1234}^{XY}(t)&=&\langle \Psi_0 | U^{xy\dagger} (t) \hat{\chi}(1234) U^{xy} (t) |\Psi_0 \rangle,
\label{xy3} \\
w_{1234}^{HS}(t)&=&\langle \Psi_0 | U^{HS\dagger} (t) \hat{\chi}(1234) U^{HS} (t) |\Psi_0 \rangle,
\label{HS3}
\end{eqnarray}
with 
$U^{xy} (t) = \exp \{ it \sum_{i<j}^{1,..,12}H^{xy}_{ij} \} $, 
$U^{HS} (t) = \exp \{ it \sum_{i<j}^{1,..,12} H^{HS}_{ij} \} $,
 and $|\Psi_0\rangle=  |s_1s_2s_3s_4 \rangle$ 
including the twelve spin qubits (Fig.~2(a)).
The number of qubits included in these calculations comes from the limitation of 
the calculation resource,
and the calculated chiralities of all the states of qubits 5 to 12 in Fig.~2(a) are 
summed and divided by $2^8$. 
We can see that when $Z_2=-Z_4$
the chirality has the finite phase. 
The $\pi/2$ phases around $t \sim 0$ 
are analyzed by the expansion 
$w_{1234}(t) \sim \langle \Psi_0| \hat{\chi}(1234) +it[\hat{\chi}(1234),H]+O(t^2)|\Psi_0\rangle$, 
with $H=\sum_{i,j=1}^{12}H_{ij}$ for $H_{ij}=H_{ij}^{HS}$ or $H_{ij}=H_{ij}^{xy}$.
Because 
$\langle \Psi_0| \hat{\chi}(1234) |\Psi_0\rangle=0$ and 
$\langle \Psi_0|[\hat{\chi}(1234),H]|\Psi_0\rangle=(1/2) Z_3 (Z_2-Z_4)$,
$w_{1234}(t)$ has a $\pi/2$ phase around $t \sim 0$.
Thus, even in the case of the always-on interaction, 
the chirality with finite phase can be obtained 
dynamically for asymmetrical spin configurations.

\begin{figure*}
\begin{center}
\includegraphics[width=12cm]{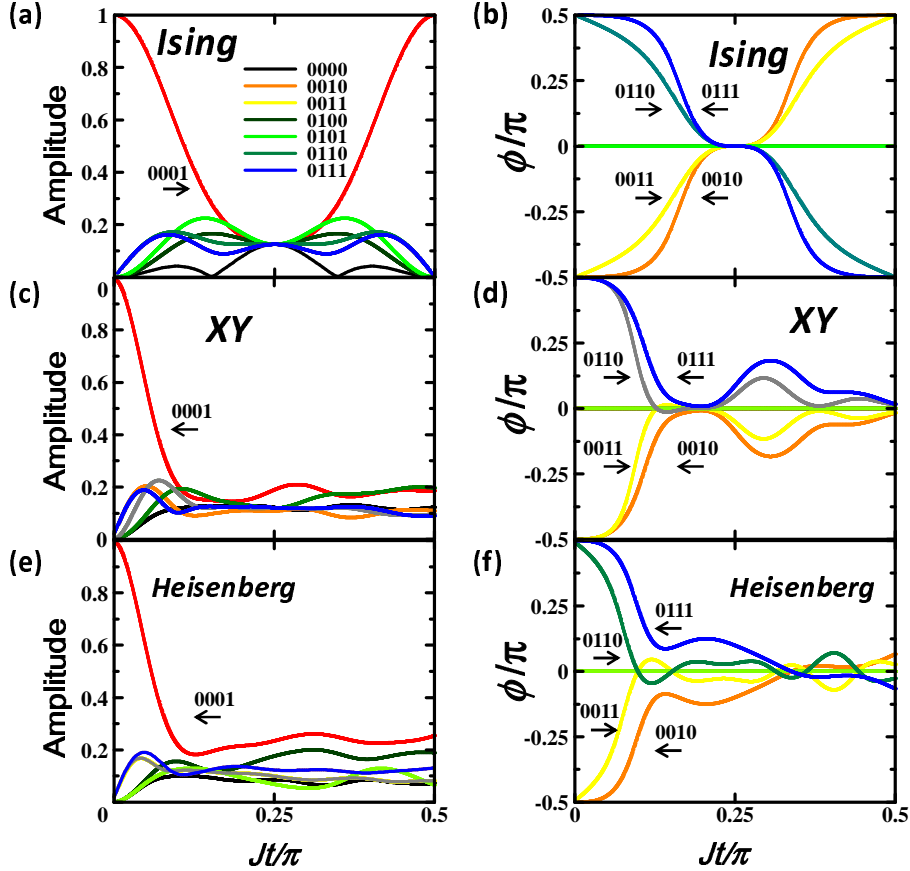}
\end{center}
\caption{
The time-dependent behavior of the chirality $w_{1234}$, Eq.~(1), 
of the square lattice. Always-on interactions are assumed for 
the configuration of spin-qubits shown in Fig.~2. 
The left and right figures show the numerical results for the amplitudes $|w_{1234}|$ 
and the phases $\phi$ of $w_{1234}=|w_{1234}|\exp i\phi$, respectively. 
(a),(b),The results for the Ising interaction calculated from the analytic form of Eq.~(\ref{XX_always}).
(c),(d),The results of the average  $w_{1234}$ for the $XY$ interaction 
numerically calculated from Eq.~(\ref{xy3}), and
(e),(f), the results of the average  $w_{1234}$ for the Heisenberg interaction
numerically calculated from Eq.~(\ref{HS3}).
For the $XY$ and the Heisenberg interactions, after obtaining $w_{1234}$ for all $2^{12}$ configurations, the average values of $w_{1234}$  are taken over the spin states 
of the site $5-12$. See Appendix \ref{sec:cal} for details.
As seen from (b), the colored patterns (0010,0011,0110,0111) in Fig.~2(b) have 
a phase  $\pi/2$ at $t\sim 0$.
}
\label{timedep}
\end{figure*}
\section{Discussions}\label{sec:discussions}
The  chiralities calculated here include no relaxation process. 
In reality, qubit systems couple to the environment and decohere.
In the case of GaAs QDs, the coherence is lost 
mainly through the interaction with the nuclear spins.
For $J\approx 0.1$--$1~\mu$eV~\cite{Petta}, 
the coherent change of the chirality is expected to be 
in the period of $(2J)^{-1} \approx  2.0$--$20$~nsec, 
which is in the range of  the experimental coherence times($<$ 50~ns~\cite{Koppens}). 

As shown in Ref.~\cite{WWZ}, there is a relationship between the 
expectation value $E_{ijk}$ and the Berry phase ${\cal B}_{ijk}$given by
${\cal B}_{123}-{\cal B}_{132}=(i/2)E_{123}$.
Thus, when the Berry phase can be detected 
as shown in Refs~\cite{Bertlmann,Filipp}, 
it might be possible to compare the calculated results here with experiments
based on Eq.(\ref{WWZ1}).

In this paper, we considered only product states as the initial states 
of the qubits. 
The time-dependent chirality of entangled states 
is interesting and important.
However,  
because there are many types of entangled states,
the chirality of entangled states 
should be discussed in a separated paper for the sake of clarity.
Even if we focus on some specific entangled states, 
there are still a lot of things to classify the results.
As an example, let us consider the chirality of the ground state of the Ising $ZZ$
interaction on a square lattice. The ground state of the four qubits 
on a square lattice is a degenerated state given by
   $ A|0101\rangle+B|1010\rangle$ with an eigenvalue of $ -4J$
($A$ and $B$ are arbitrary constants).
Then the chirality is calculated as 
$-(A^*B+AB^*)=-\sin (2p) \cos (q)$ when $A=\cos(p)$, $B=\sin(p)\exp(iq)$. 
Thus, depending on the coefficients $A$ and $B$,
the chirality changes variously.

\section{Summary}\label{sec:summary}
In summary, we have shown simple methods to generate 
the chiral spin Hamiltonian from 
conventional spin-spin interactions. 
We have also shown that, even for the always-on 
interaction (the conventional spin system), 
the chiral spin state is realized if the initial 
state is appropriately prepared.

\acknowledgements
The author would like to thank A. Nishiyama, K. Muraoka, S. Fujita, F. Nori, C. Bruder, 
H. Goto and H. Kawai for discussions.

\appendix
\renewcommand{\thefigure}{A\arabic{figure}}

\setcounter{figure}{0}
\section{Derivations of chiralities and basic relations}\label{sec:A}
The formulation of the chirality, equation (1), is derived  
by assuming the half-filled case (one spin per site). 
The electron annihilation operator $\hat{c}_{is}$ ($s \in \{ \uparrow, \downarrow\}$), 
and the Pauli matrices have the relationship given by
\begin{eqnarray}
\sigma_{i}^x &=& c_{i\uparrow}^\dagger c_{i\downarrow}
+c_{i\downarrow}^\dagger c_{i\uparrow}, 
\nonumber \\
\sigma_{i}^y &=&
=-i ( c_{i\uparrow}^\dagger c_{i\downarrow}
-c_{i\downarrow}^\dagger c_{i\uparrow}  ), \nonumber \\
\sigma_{i}^z &=&
=c_{i\uparrow}^\dagger c_{i\uparrow}
-c_{i\downarrow}^\dagger c_{i\downarrow}, \nonumber 
\end{eqnarray}
with 
$\sigma_i^p \equiv  \frac{1}{2} (\sigma^x_i + i \sigma^y_i )
= c_{i\uparrow}^\dagger c_{i\downarrow}$, 
and
$\sigma_i^m \equiv \frac{1}{2} (\sigma^x_i - i \sigma^y_i )
= c_{i\downarrow}^\dagger c_{i\uparrow}$.

The explicit form of the chirality $w_{1234}$ given by equation (1) is 
directly derived by inserting 
$\hat{\chi}_{ij} \equiv \sum_{s=\uparrow \downarrow} \hat{c}_{is}^\dagger \hat{c}_{js}$,
and we have,
\begin{eqnarray}
\hat{\chi}_{12}\hat{\chi}_{23}\hat{\chi}_{34}\hat{\chi}_{41}
&=&
 a_p  c_p
+b_p d_m
+ a_m c_m
+b_m d_p,
\label{chi1234}
\end{eqnarray}
where
\begin{eqnarray}
a_p  &=&  n_{2\downarrow}n_{3\downarrow}  +\sigma_{2}^m \sigma_{3}^p,
\nonumber \\
a_m &=& n_{2\uparrow} n_{3\uparrow}+\sigma_{2}^p\sigma_{3}^m, 
\nonumber \\
b_p &=&  -\sigma_{2}^p n_{3\downarrow} -n_{2\uparrow} \sigma_{3}^p,
\nonumber \\
b_m &=&  -\sigma_{2}^m n_{3\uparrow}  -n_{2\downarrow}\sigma_{3}^m, 
\nonumber \\
c_p&=& n_{1\uparrow}  n_{4\downarrow}  - \sigma_{1}^p \sigma_{4}^m, 
\nonumber \\
c_m&=& n_{1\downarrow}  n_{4\uparrow} -\sigma_{1}^m \sigma_{4}^p,
\nonumber \\
d_m &=& \sigma_{1}^m n_{4\downarrow} - n_{1\downarrow}  \sigma_{4}^m,
\nonumber \\
d_p &=& \sigma_{1}^p  n_{4\uparrow} - n_{1\uparrow} \sigma_{4}^p,
\nonumber
\end{eqnarray}
with 
$n_{i\uparrow}=  c_{i\uparrow}^\dagger c_{i\uparrow}$, and
$n_{i\downarrow}=c_{i\downarrow}^\dagger c_{i\downarrow}$.  

When we derive expectation values accompanying with 
the unitary transformations 
$U_{ij}^{HS}(\theta ) =\exp \{i\theta \vec{\sigma}_i \cdot \vec{\sigma}_j \}$, 
$U_{ij}^{XY}(\theta ) =\exp \{i\theta [\sigma_i^x \sigma_j^x+\sigma_i^y \sigma_j^y] \}$, 
and
$U_{ij}^{XX}(\theta ) =\exp \{i\theta \sigma_i^z \sigma_j^z \}$, 
we use the equations given by~\cite{tanaSW}
\begin{eqnarray}
& & U_{12}^{HS} (-\theta) \sigma^{z}_{1} U_{12}^{HS} (\theta)  = 
\cos^2 (2\theta) \sigma^{z}_{1}  \nonumber \\
& & +\sin^2 (2\theta) \sigma^{z}_{2} 
+\frac{1}{2} \sin (4\theta) (\sigma^{x}_{1}\sigma^{y}_{2}-\sigma^{y}_{1}\sigma^{x}_{2})\:. 
\label{Heisenberg1}
\end{eqnarray}
and its cyclic relations($x\rightarrow y \rightarrow z\rightarrow  x$),
for the Heisenberg interaction,
\begin{eqnarray}
\! U_{12}^{XY}(-\theta ) \sigma^x_{1} U_{12}^{XY}(\theta ) \!\!\!&\!=\!&\!\! \cos (2\theta) \sigma^x_{1} -\sin (2\theta) \sigma^z_{1} \sigma^y_{2}
\label{XYa}, \\
\! U_{12}^{XY}(-\theta ) \sigma^y_{1} U_{12}^{XY}(\theta ) \!\!\!&\!=\!&\!\! \cos (2\theta) \sigma^y_{1} +\sin (2\theta) \sigma^z_{1}\sigma^x_{2} 
\label{XYb}, \\
\! U_{12}^{XY}(-\theta )\sigma^z_{1} U_{12}^{XY}(\theta ) \!\!\!&\!=\!&\!\! \cos^2 (2\theta) \sigma^z_{1} +\sin^2 (2\theta) \sigma^z_{2}  \nonumber\\
\!\!\!&\!+\!&\!\!\frac{1}{2}\sin (4\theta)
[\sigma^x_{1}\sigma^y_{2}-\sigma^y_{1}\sigma^x_{2}]\:, 
\label{XYc} 
\end{eqnarray}
for $XY$ interaction, 
and
\begin{eqnarray}
& & U_{12}^{XX}(-\theta ) \sigma^y_{1} U_{12}^{XX}(\theta ) 
\!=\! \cos (2\theta) \sigma^y_{1} +\sin (2\theta) \sigma^z_{1} \sigma^x_{2},
\label{ZZa}\\
& & U_{12}^{XX}(-\theta ) \sigma^z_{1}U_{12}^{XX}(\theta ) 
\!=\! \cos (2\theta) \sigma^z_{1} -\sin (2\theta) \sigma^y_{1} \sigma^x_{2}\:,
\label{ZZb}
\end{eqnarray}
for Ising interaction.
The expectation values 
$\langle \Psi_0 | \hat{\chi}_{12}\hat{\chi}_{23}\hat{\chi}_{34}\hat{\chi}_{41} |\Psi_0 \rangle$ are
estimated by the product states $ |\Psi_0 \rangle =|s_1s_2s_3s_4 \rangle$ 
($s_i \in \{\uparrow, \downarrow\}$). 

\section{General form of the left method of Table I}\label{sec:tableIa}
The general form for the Heisenberg interaction is given by
\begin{eqnarray}
w_{1234}^{HS} &=&
\frac{1}{8} \left\{
1-Z_1Z_2Z_3Z_4+Z_4Z_B-Z_1Z_A \right. \nonumber \\
&+& \left. e^{4iJt} (Z_2Z_3-Z_1[Z_B+Z_4]+Z_4Z_A \right. \nonumber \\
&+& \left. i \sin (4Jt)[ Z_4-Z_1][Z_2-Z_3] )\right\},
\end{eqnarray}
where
$Z_A=\cos^2(2Jt) Z_2 +\sin^2 (2Jt) Z_3$ and $Z_B=\sin^2(2Jt) Z_2 +\cos^2 (2Jt) Z_3$.
For the $XY$ interaction, we have
\begin{eqnarray}
w_{1234}^{XY} &=&
\frac{1}{8} \{
1-Z_1Z_2Z_3Z_4+Z_4Z_B-Z_1Z_A \nonumber \\
&+& e^{4iJt} (Z_2Z_3-Z_1[Z_B+Z_4]+Z_4Z_A ) \nonumber \\
&+& i e^{2iJt} \sin (4Jt)[ Z_4-Z_1][Z_2-Z_3] )\},
\end{eqnarray}
For the Ising $XX$ interaction, we have
\begin{eqnarray}
w_{1234}^{XX} &=&
\frac{1}{8} \{
(1-e^{2iJt}Z_1Z_2)(1+e^{-2iJt}Z_3Z_4) \nonumber \\
&+&(-Z_1+e^{2iJt}Z_2)(Z_3+e^{2iJt}Z_4) 
\}.
\end{eqnarray}

\section{General form of the right method of Table I}\label{sec:tableIb}
The general expression of the chirality of the right method of Table I 
is 
given by  
$w_{1234}^{XX}=G_{23}^-(\theta) H_{14}^+(\theta)
+G_{23}^+(\theta) H_{14}^-(\theta)$ for the Ising $XX$ interaction, 
and the 
$XY$ and Heisenberg cases provide the same form 
of
$w_{1234}^{XY}=w_{1234}^{hs}=F_{23}^-(\theta) H_{14}^+(2\theta)
+F_{23}^+(\theta) H_{14}^-(2\theta)$.
where
\begin{eqnarray}
F_{ij}^\pm(\theta) &\equiv& ( [1\pm Z_i][1\pm Z_j]\pm i\sin{4\theta}[Z_i-Z_j])/4,   \\
G_{ij}^\pm(\theta) &\equiv& (1\pm Z_i e^{2i\theta}) (1\pm Z_j e^{-2i\theta})/4, \\
H_{ij}^\pm(\theta) &\equiv& (1-Z_i Z_j\pm  e^{-2i\theta}[Z_i-Z_j])/4, 
\end{eqnarray}
Thus, in order to obtain a finite phase, $Z_4=-Z_1$ or $Z_3=-Z_2$ are 
necessary. 
Table I shows the results for the simple case of  $Z_4=-Z_3=Z_2=-Z_1$.

\section{Numerical calculations}\label{sec:cal}
In Fig.3c-f, we have directly calculated equation (1) for the $XY$ and the Heisenberg 
interactions, as expressed by Eq.(\ref{xy3}) and (\ref{HS3}), respectively.
There are $2^{12}$ patterns of the spin configurations in Fig.~2a. 
Depending on the spin configuration over the twelve sites of Fig.2a,
the time-dependent chirality changes in various ways. 
Fig.~4 shows a sample of the results of the Heisenberg interaction
of the pattern 0010 of Fig.2b.
`2,18,34,50,66,82' correspond to the spin configurations of the 12 sites.
The spin configuration can be expressed by a binary form of  
`$i_{12}i_{11}i_{10}i_{9}~i_{8}i_{7}i_{6}i_{5}~i_{4}i_{3}i_{2}i_{1}$',
such that $i_j=0$ or 1 ($j=1,...,12$) depending on spin-up or spin-down, respectively.
Then the binary form can be transformed to the decimal number 
given by $i_{12}\cdot 2^{12}+i_{11}\cdot 2^{11}+i_{10}\cdot 2^{10}+
i_{9}\cdot 2^9+i_{8}\cdot 2^8+i_{7}\cdot 2^7+i_{6}\cdot 2^6+i_{5}\cdot 2^5+i_{4}\cdot 2^4+i_{3}\cdot 2^3+i_{2}\cdot 2^2+i_{1}$.
For example, `2' corresponds to '0000~0000~0010', which means that 
the spin of site 2 is flipped, and 
'18' corresponds to 0000~0001~0010, which means that 
the spins of sites 2 and  5 are flipped.
It is seen that the phase of the chirality around $t\sim0$ is $\pi/2$ for all 
configurations. The time-dependent behaviors for $t>0$ are
different depending on their configuration. 
The results shown in Fig.~3 are the averaged results over 
 all the configurations.

\begin{figure*}
\begin{center}
\includegraphics[width=15cm]{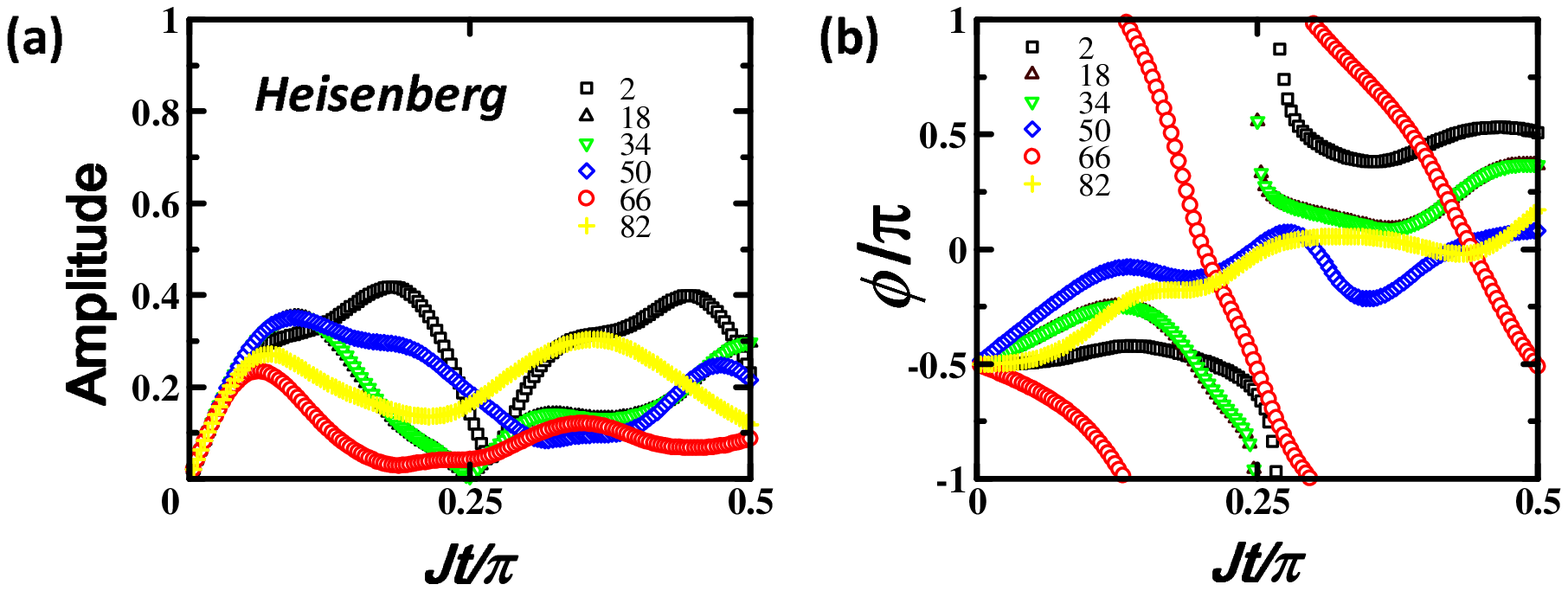}
\end{center}
\caption{
Examples of the calculation of the time-dependent chirality 
for the Heisenberg Hamiltonian
of the twelve qubits before averaging.
The `2', `18', `34', `50', `66', and `82' express  
the qubit states over the twelve sites when 0=$\uparrow$ and 1=$\downarrow$
such as
$2=0000~0000~0010$,
$18=0000~0001~0010$,
$34=0000~0010~0010$,
$50=0000~0011~0010$,
$66=0000~0100~0010$, and
$82=0000~0101~0010$.
The last four digits `0010' corresponds to the configuration of 
the pattern `0010' of Fig.~2b.
Other states in the $2^8$ configurations show similar behaviors.
Fig.~3(e)(f) show the average of these results.
} 
\label{triangle}
\end{figure*}


\end{document}